\documentclass[aps,twocolumn,groupedaddress,superscriptaddress,floatfix,pra]{revtex4-1}
\usepackage{amsmath}
\usepackage{mathtools}
\usepackage{amssymb}
\usepackage{graphicx}
\usepackage{textcomp}
\usepackage{bm}	
\usepackage{soul}
\usepackage{cancel}
\usepackage{ulem}

\usepackage[usenames,dvipsnames]{color}

\usepackage[braket, qm]{qcircuit}

\begin{document}

\title{Charging a quantum battery via non equilibrium heat current}

\author{Francesco Tacchino}
\altaffiliation[Present address:]{
 \textit{IBM Quantum, IBM Research -- Zurich, S\"{a}umerstrasse 4, CH-8803 R\"{u}schlikon, Switzerland}}
\affiliation{Dipartimento di Fisica, Universit\`a di Pavia, via Bassi 6, I-27100, Pavia, Italy}

\author{Tiago F. F. Santos}
\affiliation{Instituto de F\'isica, Universidade Federal do Rio de Janeiro, CP68528, Rio de Janeiro, Rio de Janeiro 21941-972, Brazil}

\author{Dario Gerace}
\affiliation{Dipartimento di Fisica, Universit\`a di Pavia, via Bassi 6, I-27100, Pavia, Italy}

\author{Michele Campisi}
\affiliation{NEST, Istituto Nanoscienze-CNR and Scuola Normale Superiore
P.zza San Silvestro 12, I-56127 Pisa, Italy}
\affiliation{Dipartmento di Fisica e Astronomia, Universit\`a di Firenze, via Sansone 1, I-50019, Sesto Fiorentino (FI), Italy}
\affiliation{INFN - Sezione di Pisa, Largo Bruno Pontecorvo 3, I-56127 Pisa, Italy}

\author{Marcelo F. Santos}\email{mfsantos@if.ufrj.br}
\affiliation{Instituto de F\'isica, Universidade Federal do Rio de Janeiro, CP68528, Rio de Janeiro, Rio de Janeiro 21941-972, Brazil}

\pacs{xxxx, xxxx, xxxx}                           
\date{\today}
\begin{abstract}
When a quantum system is subject to a thermal gradient it may sustain a steady non-equilibrium heat current, by entering into a so-called non equilibrium steady state (NESS). Here we show that NESS constitute a thermodynamic resource that can be exploited to charge a quantum battery. This adds to the list of recently reported sources available at the nano-scale, such as coherence, entanglement and quantum measurements. We elucidate this concept by showing analytic and numerical studies of a two-qubits quantum battery that is alternatively charged by a an incoherent heat flow and discharged by application of a properly chosen unitary gate. The presence of a NESS for the charging step guarantees steady operation with positive power output. Decreasing the duration of the charging step results in a time periodic steady state accompanied by increased efficiency and output power. The device is amenable to implementation with different nanotechnology platforms.
\end{abstract}

\maketitle

With the fast advancement of Quantum Technologies, understanding and mastering the microscopic mechanisms that govern energy exchanges in nanoscale systems and devices has entered in the limelight of current research. This has led to a flourishing research activity aimed at singling out novel approaches to implement nano-thermal machines that take advantage of peculiar resources, e.g., of quantum nature, which are typically available only at minute scales and low-temperatures. It has already been pointed out that quantum coherences represent a genuinely quantum resource to be employed for fuelling heat engines \cite{Scully03Science299,Klatzow19PRL122}. A nano-heat engine operating thanks to a purely quantum effect (the AC Josephson effect) has been recently proposed \cite{Marchegiani16PRAPP6}, while it is expected that entanglement may, under special circumstances, constitute a thermodynamic resource as well~\cite{PhysRevLett.118.150601, NJP3_12_2017, CorAsRes1,tacchino_steady_2018}. Recently, another genuinely quantum effect, namely the invasive nature of the quantum measurement process, has been shown to also constitute a fuel for quantum heat engines~\cite{Buffoni19PRL122}. Other works have focused on further aspects such as back-flow of information from the heat reservoirs \cite{Bylicka15SR16}, feedback control \cite{Campisi17NJP19,Quan06PRL97,Toyabe10NP6}, and non-thermal-equilibrium dynamics
~\cite{PhysRevLett.122.210601,Nicole}, among others (for a recent review, see Ref.~\onlinecite{Kosloff14ARPC65}).

\begin{figure}[b]
    \centering
    \includegraphics[width=0.8\columnwidth]{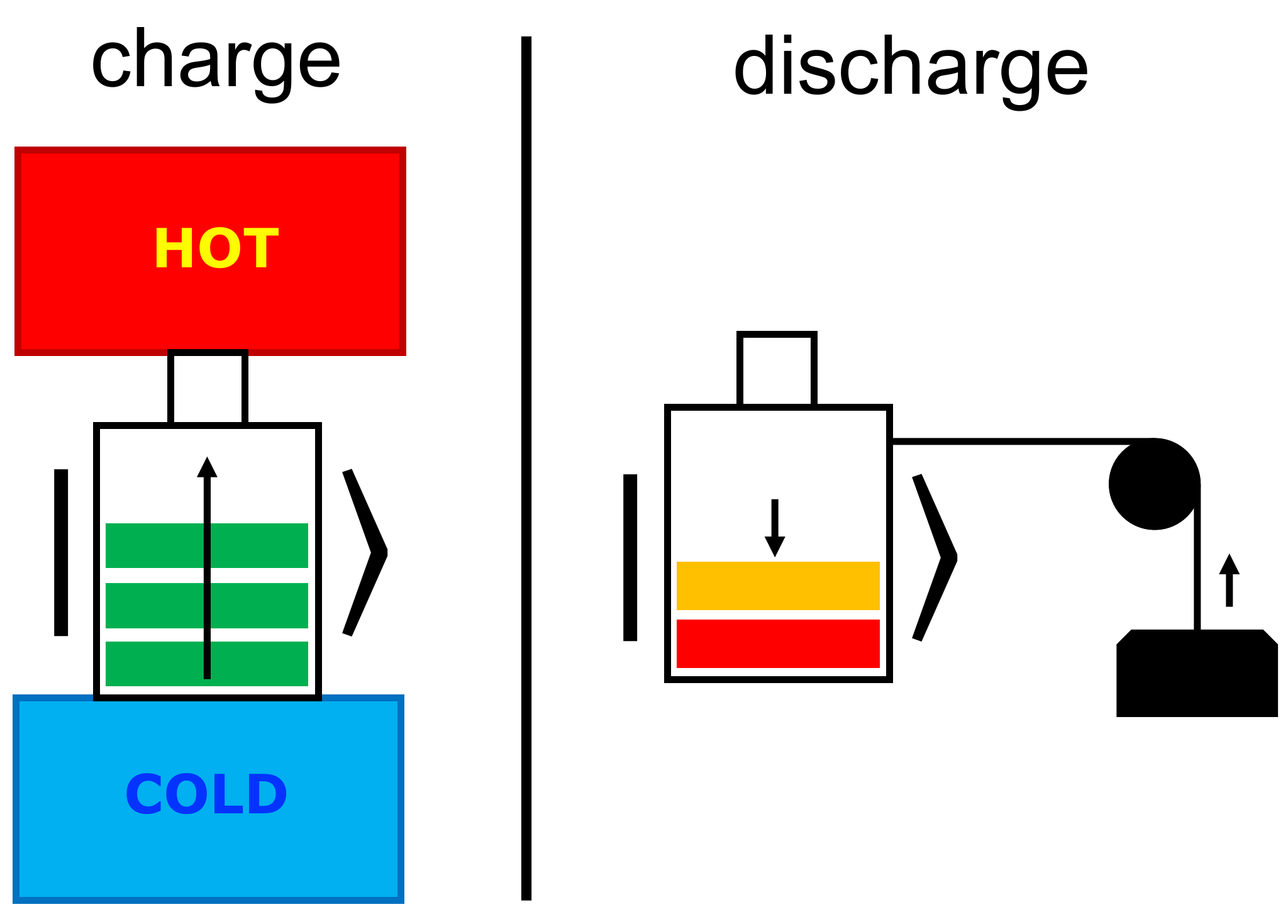}
    \caption{Two stroke NESS-based quantum heat engine scheme: first the quantum working fluid playing the role of a quantum battery is charged, as it is traversed by an incoherent heat flux, and reaches a steady state; then, energy is extracted in the form of work by suitably coupling the battery to a work source that evolves it unitarily.
    }
    \label{Figure1}
\end{figure}

Within this research framework, a problem that has recently become increasingly relevant is how to design and implement the charging of quantum batteries~\cite{PhysRevLett.118.150601, Binder15NJP17, Ferraro18PRL120, Ferraro20EPJWC230, Andolina1, Andolina2}. These are quantum systems with a finite spectrum, which can be manipulated in such a way as to store and deliver energy to external agents. So far, the research has focused on charging a quantum battery by means of unitary operations \cite{PhysRevLett.118.150601,Binder15NJP17}, coupling to other quantum systems \cite{Ferraro18PRL120}, as well as performing projective measurements \cite{Solfanelli19JSM}. Particularly interesting is the charging of a quantum battery by means of thermal mechanisms, which would require much less control and greater stability \cite{Hovhannisyan20PRRES2}. However, (weak) coupling to a thermal bath is not effective, as it results in reaching an uncharged Gibbs state (i.e., a ``passive'' one)~\cite{Pusz78CMP58, Lenard78JSP19}. On the other hand, Hovhannisyan \textit{et al.} have shown that strong coupling to a thermal bath may result in a robust active state \cite{Hovhannisyan20PRRES2}, and Guarnieri \textit{et al.} \cite{PhysRevLett.121.070401} demonstrated that steady state energy coherences can be generated in a two-level system interacting with a single heat bath over a wide range of coupling regimes. It has also long been known that the interaction with two thermal baths can induce some form of activity~\cite{PhysRevLett.2.262}. More generally, interaction with two or more baths may induce non-equilibrium distributions of populations in composite systems, and possibly coherence in their energy eigenbasis \cite{PhysRevE.76.031115,Mitchison_2018,PhysRevB.97.035432}, potentially producing, under the proper circumstances, active states.

In this work, we show that a quantum battery can be charged when it is traversed by an incoherent heat flow resulting from the weak coupling of a quantum system to two thermal baths at different temperatures. Here, the charging is caused by the application of a thermal gradient, while for common batteries it would be driven by the application of an electrochemical potential gradient. A quantum thermal machine can accordingly be realized as a two-stroke engine (see. Fig. 1), where in the first stage a non-equilibrium steady state (NESS) develops in time, thus leading the quantum system to an active state (namely a state from which one can extract energy by means of a unitary operation); while in the second stage, a unitary operation allows to extract the highest possible amount of energy stored in the quantum state of the system (i.e., the so called \textit{ergotropy}~\cite{allahverdyan_maximal_2004}). The quantum battery is now in its \textit{passive} state associated to the NESS, ready to be recharged and then start the cycle again. Our protocol thus realizes a complete thermodynamic cycle fully based on a non-equilibrium and non-Gibbsian active state generated by the weak interaction with two thermal baths.

We propose an implementation of the device with state-of-art quantum optics technology, the working system being two coupled qubits. Numerical analysis validates the working principle and gives quantitative estimates for the expected efficiency and power output, respectively. We further analyse the thermodynamic performance as a function of the duration of the charging stroke. Remarkably, we show that for short charging times the quantum engine enters into a time periodic Floquet-type NESS after a few cycles, which is accompanied by enhanced output power and efficiency.

\section{Theoretical background}

In the following we shall adopt the units convention where $k_B=1, \hbar=1$.

During the charging stroke the quantum system evolves while in contact with two heat baths at different temperatures. We shall model its dynamics via a Lindblad master equation
\begin{align}
    \dot{\rho} = \mathcal{\bar L}(\rho)= -i[H,\rho] + \mathcal{L}(\rho) \, .
\label{eq1}
\end{align}
The Lindblad superoperator, $\mathcal{\bar L}$, contains a unitary part, stemming from the free evolution of the system dictated by its Hamiltonian, $H$, and a dissipative part, $\mathcal{L}=\sum_j L_j$, accounting for the coupling to the baths. Here the Lindblad operators $L_j$ are of the form
$L_j(\rho) = \Gamma_j[J_j\rho J_j^\dagger -\frac{1}{2}\{J_j^\dagger J_j, \rho \}]$,
where $\{A,B\}=AB+BA$, $J_j$ are jump operators, and $\Gamma_j$ are the related jump rates \cite{Breuer-Petruccione02}. Formally, the solution of Eq. (\ref{eq1}) reads
\begin{equation}
\rho(\tau) = e^{\mathcal{\bar{L}}\tau} \rho= \sum_k M_k(\tau) \rho M_k(\tau)^\dagger
\label{eq:solutionLindblad}
\end{equation}
where the Kraus operators, $M_k(\tau)$, obey the relation $\sum_k M_k^\dagger(\tau) M_k(\tau)=1$, depend on the duration $\tau$ of the charging stroke, on the bath temperatures, and on the details of the couplings to the baths. 
We assume that the Lindblad operator has a fixed point,  $\rho_{NESS}$, representing a non-equilibrium steady state, $\mathcal{\bar L}(\rho_{NESS}=0)$, which the system approaches for times longer than a characteristic time scale, $\tau_{NESS}$.

In the discharging stage a unitary operation, $U_{\rho(\tau)}$, is applied on the quantum system by means of the switching on of a properly designed coupling with an external work source (e.g., an electromagnetic field). The underlying assumption here is that the duration of this operation, $\tau_d$, is such that $\tau_d \ll \tau$. We shall accordingly treat this step as instantaneous, hence $\tau$ will denote the duration of a complete cycle. Given the state $\rho(\tau)$ at the beginning of the discharging stage, among all possible unitary operations, we obviously choose the one that extracts the maximal amount of energy. As shown in Ref.~\onlinecite{allahverdyan_maximal_2004}, such operation reads 
\begin{equation}
\mathcal{E}_{\rho(\tau)}= \sum_{k,j} r_k E_j (|\langle r_k|E_j \rangle|^2 - \delta_{kj}) \,
\end{equation}
where $E_j$ are the eigenenergies of system Hamiltonian $H$ ordered in increasing magnitude, i.e., $E_i \geq E_j$ for $i>j$, and $r_k$ are the eigenvalues of the density operator $\rho(\tau)$ in decreasing order, i.e., $r_i \leq r_j$ for $i>j$. Denoting their respective eigenvectors as $|E_j\rangle$ and $|r_j\rangle$, the related unitary operation reads $U_{\rho(\tau)}=\sum |E_j\rangle \langle r_j|$, where we explicitly denoted its dependence on $\rho(\tau)$. The quantity $\mathcal{E}_{\rho(\tau)}$ is known as the ergotropy of state $\rho(\tau)$ \cite{allahverdyan_maximal_2004}.
Note that when $\rho(\tau)$ and $H$ commute, the unitary operation amounts to a permutation of the occupation probabilities of each energy eigenstate. The application of the unitary $U_{\rho(\tau)}$ leaves the system in a so called passive state: a state from which one cannot extract any energy by means of unitary operations. In order for our engine to output work in a steady manner, it is crucial that the state at the beginning of each discharging stage be active, namely we need that the interaction with the baths activate the quantum system. In this respect, it is worth stressing that (weak) interaction with a single bath cannot fulfil this condition since thermalisation leads to the passive Gibbs state: in order to activate a system by means of (weak) thermal interactions, two baths are required at least. 
We also remark that in the case of a battery having only two levels, coupling to two thermal baths will result in a NESS that is in Gibbs form, $e^{- H/T^*}/Z^* $ for some $T_1\leq T^*\leq T_2 $, namely a passive state. Therefore, a battery having more than two levels is also a necessary ingredient.

Overall the (completely positive and trace preserving) map that advances the density operator of the working substance by one cycle reads:
\begin{equation} 
\rho \rightarrow \sum_k U_{\rho(\tau)} M_k(\tau)\rho
M_k^\dagger(\tau) U_{\rho(\tau)}^\dagger \, ,
\label{map}
\end{equation}
For the engine to be able to work cyclically it is necessary that the map above has a fixed point, which we shall refer to as the operational steady state $\rho_{OSS}$. In that case, as we shall see in our proposed implementation below, after a sufficient number of cycles (which may vary depending on $\tau$) the system enters into a time-periodic Floquet-type steady state with period $\tau$. In that situation the energy gained during the charging stage balances with that given away during the discharging stage. In other words, at steady operation, the first stage pumps heat into the machine and takes the system from passive state $\rho_{OSS}$ to its active counterpart $\rho_{OSS}(\tau)$, whereas the second stage extracts energy from the machine in the form of work and resets the system back to $\rho_{OSS}$

Once the operational steady state is achieved, the power delivered reads $P = \mathcal{E}_{\rho_{OSS}(\tau)}/\tau$. The heat absorbed from the hot bath in a cycle reads \cite{Alicki79JPA12}
\begin{align}
    Q^{H}(\tau) = \int_{0}^{\tau} dt \operatorname{Tr}
    \left[ L_H(\rho_{OSS}(t))H\right] \, ,
    \label{eq:QH}    
\end{align}
where $L_H$ is the superoperator accounting for the effect of the coupling to the hot thermal source.
The thermodynamic efficiency then reads $\eta=\mathcal{E}_{\rho_{OSS}}(\tau)/Q^{H}(\tau)$.

\section{Two coupled qubits as the working fluid}
\begin{figure}
    \centering
    \includegraphics[width=\columnwidth]{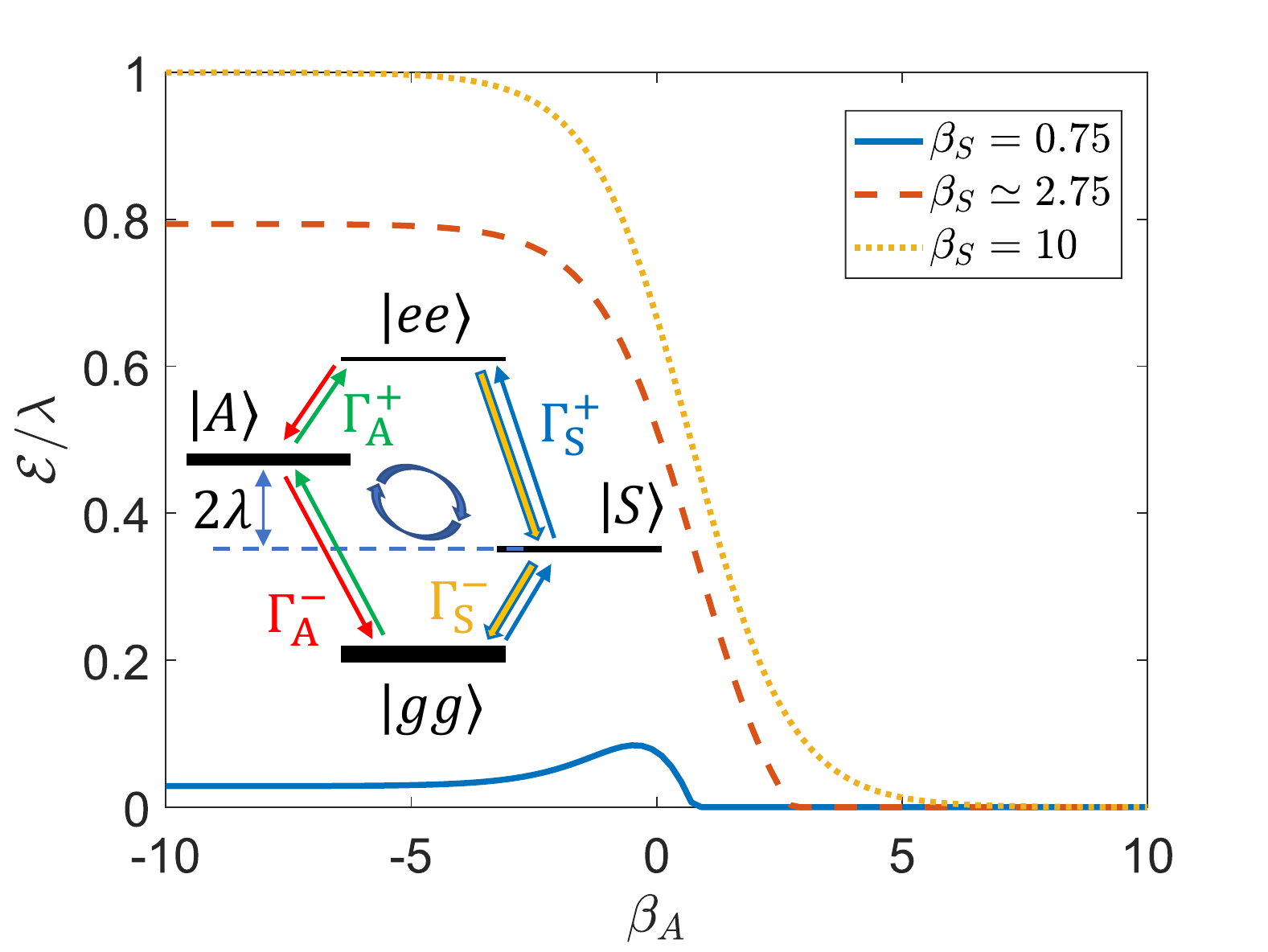}
    \caption{Steady state ergotropy of the two coupled qubits setup as a function of inverse effective temperatures. $\beta_S \simeq 2.75 $ is equivalent to $\Gamma_S^+ = \Gamma_A^+ = 3.5\cdot 10^{-4}\omega_0$, $\Gamma_S^- = 5.5\cdot 10^{-3}\omega_0$ and $\Gamma_A^- = 5\cdot 10^{-4}\omega_0$. These values were used in other simulations reported below. The inset shows the configuration of the energy levels with the transitions induced by external pumping and dissipation channels. The thickness of the levels represents the typical population distribution required to operate the thermal engine via the $|A\rangle \leftrightarrow |S\rangle$ exchange.}
    \label{fig:model}
\end{figure}

From now on, we consider that the working fluid consists of two degenerate and mutually coupled qubits, with energy gap $\omega_0$ and coupling strength $\lambda \ll \omega_0$, schematically represented in Fig.~\ref{fig:model}a. The Hamiltonian reads 
\begin{equation}
    H = \omega_0(\sigma_+^{(1)}\sigma_-^{(1)} + \sigma_+^{(2)}\sigma_-^{(2)}) - \lambda (\sigma_+^{(1)}\sigma_-^{(2)} + h.c.)
\end{equation}
where $2\sigma_\pm = \sigma_x \pm i\sigma_y$ and $\{\sigma_j\}$ for $j = x,y,z$ are the Pauli matrices. The resulting energy levels, which can be obtained by the direct diagonalization of $H$, are shown in the inset of Fig.~\ref{fig:model} and correspond to the following set of total spin eigenstates
\begin{equation}
    \begin{cases}
        |gg\rangle \qquad & E_{gg} = 0 \\
        |S\rangle = \frac{1}{\sqrt{2}}(|ge\rangle+|eg\rangle) \qquad & E_{S} = \omega_0 - \lambda \\
        |A\rangle = \frac{1}{\sqrt{2}}(|ge\rangle-|eg\rangle) \qquad & E_{A} = \omega_0 + \lambda \\
        |ee\rangle \qquad & E_{ee} = 2\omega_0
    \end{cases}
\label{eq:eigenstates}
\end{equation}

Here $|g\rangle$ ($|e\rangle$) denotes a qubit in the ground (excited) state. The two qubits are also coupled to two baths at temperatures $\mathcal{T}_S$ and $\mathcal{T}_A$, each acting respectively on the symmetric (S) and anti-symmetric (A) transitions of the system, as depicted in Fig.~\ref{fig:model}. The overall dynamics of the machine will then follow Eq.~(\ref{eq1}) with $\mathcal{L}=L_A+L_S$ where
\begin{eqnarray}
    L_A(\rho) & = & \Gamma^+_A [A^\dagger \rho A-\frac{1}{2}\{AA^{\dagger},\rho \}]+\Gamma^-_A [A \rho A^\dagger-\frac{1}{2}\{A^{\dagger}A,\rho \}],\nonumber \\
    L_S(\rho) & = & \Gamma^+_S [S^\dagger \rho S-\frac{1}{2}\{SS^{\dagger},\rho \}]+\Gamma^-_S [S \rho S^\dagger-\frac{1}{2}\{S^{\dagger}S,\rho \}] \, .
\end{eqnarray}
The rates $\Gamma^{\pm}_{A,S}$ are given by
\begin{equation}
\Gamma^+_{A,S} = \Gamma^0_{A,S}\frac{1}{e^{\omega_0/\mathcal{T}_{A,S}}-1}
\textrm{  and  } \Gamma^-_{A,S} = \Gamma^0_{A,S}\frac{e^{\omega_0/\mathcal{T}_{A,S}}}{e^{\omega_0/\mathcal{T}_{A,S}}-1},
\end{equation}
where $S^\dagger=\sigma^{(1)}_++\sigma^{(2)}_+$ and $A^\dagger=\sigma^{(1)}_+-\sigma^{(2)}_+$. Note that these expressions are valid as long as $\lambda \ll \omega_0$. A possible quantum optical scheme to produce such dynamics was introduced in Ref.~\onlinecite{tacchino_steady_2018} and will be discussed in more detail later.

\begin{figure}[t]
    \centering
 	\includegraphics[width=\columnwidth]{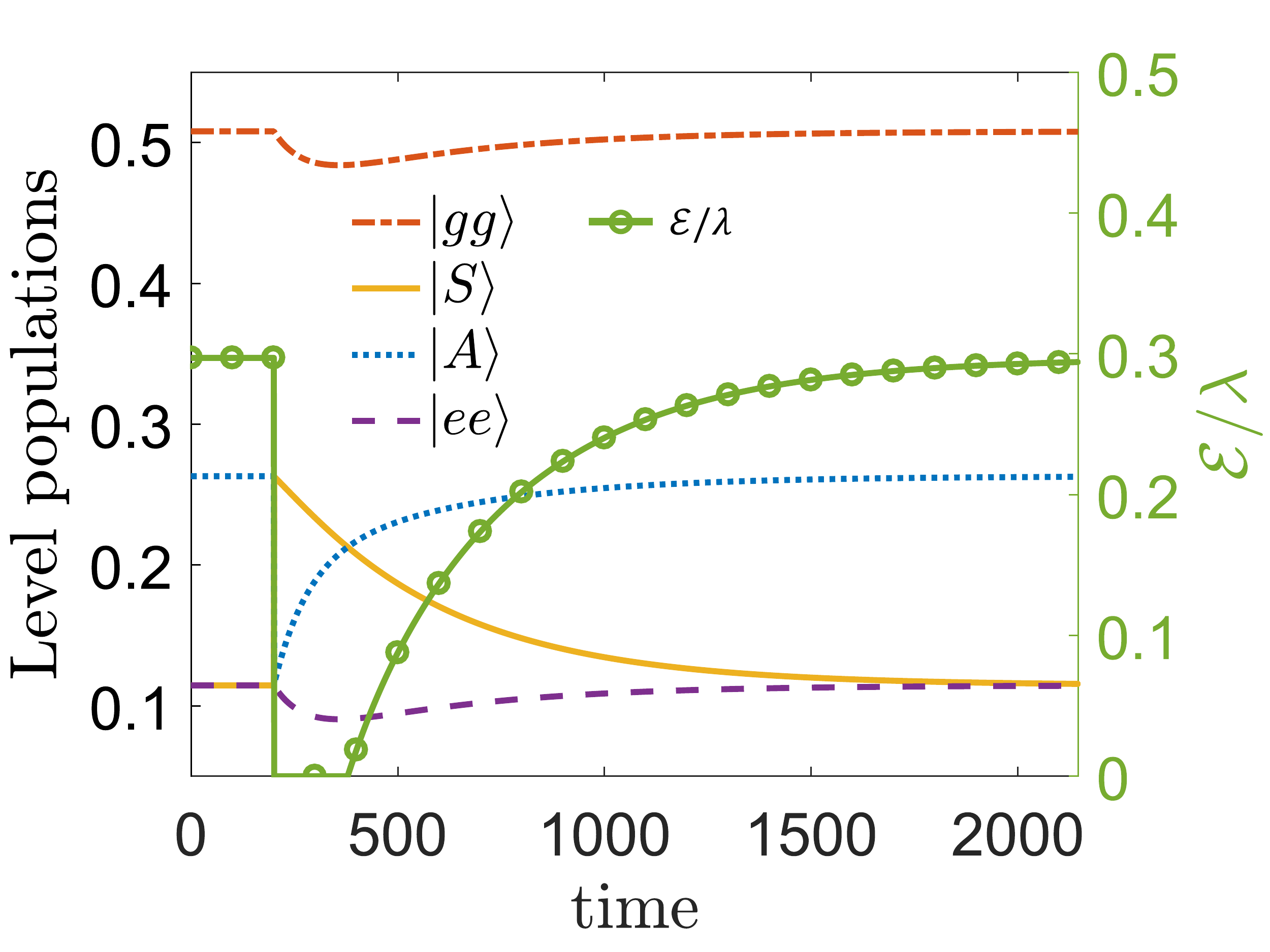}

    \caption{Full cycle for the two-qubit thermal machine with instantaneous work extraction happening after a short idle phase and complete recharging to the NESS, shown in terms of the level populations as a function of time. Here parameters are chosen as $\lambda = 10^{-2}\omega_0$, $\Gamma_k^+ = \Gamma_0 \bar{n}_k$, $\Gamma_k^- = \Gamma_0 (\bar{n}_k+1)$, with $\bar{n}_k = 1/(e^{\omega_0/\mathcal{T}_k}-1)$, $\mathcal{T}_A = 2\omega_0$, $\mathcal{T}_S = 0.1\omega_0$ and $\Gamma_0 = 10^{-3}\omega_0$.}
    \label{populations}
\end{figure}

The general expression for the heat flux from the hot source ($\mathcal{T}_A$) is given by 
\begin{eqnarray}
\dot{Q}^{A} =4\omega_{0} (\Gamma^+_A r_A - \Gamma^-_A r_e) +(\omega_{0}+\lambda)\dot{r}_A
\end{eqnarray}
which, in the limit $\tau \rightarrow \infty$ ($\dot{r}_j=0$) becomes
\begin{eqnarray}
\dot{Q}^{A} &=& 4\omega_{0} \frac{\Gamma^+_A\Gamma^-_S-\Gamma^-_A\Gamma^+_S}{\Gamma^+_A+\Gamma^+_S}r_{S_{NESS}}.
\end{eqnarray}
Note that, if $\Gamma^0_S=\Gamma^0_A$, $\dot{Q}^A$ is positive as long as $\frac{\bar{n}_A}{\bar{n}_A+1}>\frac{\bar{n}_S}{\bar{n}_S+1}$ (where $\frac{\bar{n}_k}{\bar{n}_k+1}=e^{-E_k/T_k}$), i.e. as long as there is a positive gradient temperature from $\mathcal{T}_A$ to $\mathcal{T}_S$. This heat flow unbalances the population of the intermediate levels of the system in favour of the anti-symmetric state $|A\rangle$. This creates the necessary condition for the battery to store ergotropy.  Moreover, the cyclic operation of the machine goes through operational states that are diagonal in the energy eigenbasis of the coupled qubits, i.e. of the type $\rho_{OSS} = \sum_j r_j|j\rangle \langle j|$, where $j=\{gg,S,A,ee\}$. In fact, there is a minimum period, $\tau_{min}$, beyond which a population inversion between the states A and S occurs:
\begin{equation}
    r_{gg} > r_A > r_S > r_{ee}
\label{eq:ordering}
\end{equation} 
accompanied by a non-zero ergotropy
\begin{equation}
    \mathcal{E} = 2\lambda [r_A(\tau) - r_S(\tau)],
    \label{eq:ness_ergo}
\end{equation}
where $\tau>\tau_{min}$. 
Notice that $\tau_{min}$ has to be much larger than both the correlation time of the reservoirs (for the Markovian approximation of Eq.~\ref{eq1} to hold true) and the energy extraction stage duraion  $\tau_d$ (for it to be considered adiabatic in the thermodynamic sense). However, in many applications $\tau_{min}$ can still be small enough for the cycle to operate in the ``quantum jumps'' limit $\sum_k \Gamma^{\pm}_j \tau_{min} \ll 1$. We also stress that when $\mathcal{T}_S=\mathcal{T}_A$, then $r_S=r_A$ and the ergotropy vanishes, which makes it clear that a temperature gradient is indeed crucial for the engine to work, as expected.

The machine here presented has some very distinctive properties. First of all, it is fuelled by a non-equilibrium steady state. Second, for the particular working fluid analysed, $\rho_{OSS}$ is, in general, entangled~\cite{tacchino_steady_2018} and there are three equally good unitary operations that maximize the extracted work: either a two-qubit operation $U_{\mathcal{E}_G}=e^{-i(\sigma_z^{(1)}-\sigma_z^{(2)})\frac{\pi}{4}}$ or local ones  $U_{\mathcal{E}_{i}} = e
^{-i\sigma_z^{(i)}\frac{\pi}{2}}$ where, $i=\{1,2\}$ indicates the qubit where $U_{\mathcal{E}_i}$ is applied.  That means that work can be fully extracted from one side or half of it simultaneously from both sides.

\begin{figure}[t]
    \centering
    \includegraphics[width=\columnwidth]{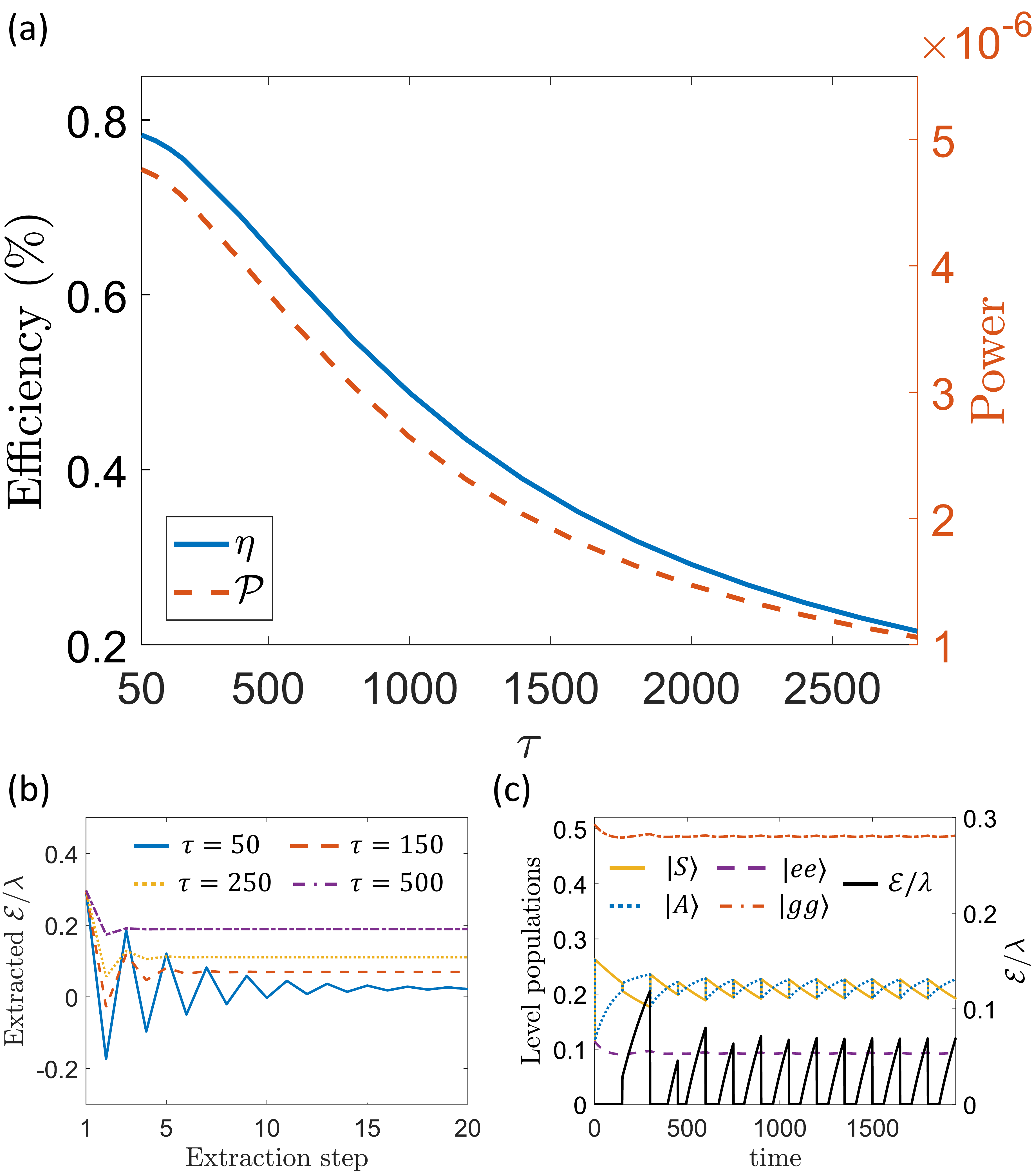}
    \caption{(a) Efficiency and power for the thermal machine operated at $\rho_{OSS}$ as a function of $\tau$. (b) ergotropy extracted per step for different choices of $\tau$ starting from $\rho_{NESS}$ and converging to $\rho_{OSS}$; (c) dynamics of the system reaching the operational steady state for  $\tau = 150$. In all panels we set $\lambda = 10^{-2}\omega_0$, $\Gamma_k^+ = \Gamma_0 \bar{n}_k$, $\Gamma_k^- = \Gamma_0 (\bar{n}_k+1)$, with $\bar{n}_k = 1/(e^{\omega_0/\mathcal{T}_k}-1)$, $\mathcal{T}_A = 2\omega_0$, $\mathcal{T}_S = 0.1\omega_0$ and $\Gamma_0 = 10^{-3}\omega_0$.}
    \label{efficiency}
\end{figure}

Third the efficiency of the machine depends on the chosen $\rho_{OSS}$. In Fig.~\ref{populations}, we show a full recharging stage as a function of time. At $t=0$ the system is in $\rho_{NESS}$ when $U_\mathcal{E}$ swaps the $|A\rangle$ and $|S\rangle$ populations. Then, the heat flowing from bath $\mathcal{T}_A$ to $\mathcal{T}_S$ recharges the battery. Note that even though larger periods allow for more stored energy in $\rho_{OSS}$, the most efficient operation takes place for shorter cycles . This is confirmed in Fig.~\ref{efficiency} where we plot the efficiency and power as a function of the period of the cycle. Efficiency reaches a maximum plateau in the short cycle limit, when $\sum_j \Gamma^{\pm}_j\tau \ll 1$. In this case, up to first order in $\tau$, both the ergotropy of $\rho_{OSS}$,
\begin{equation}
\mathcal{E}_\tau =4\lambda \tau K(\Gamma^-_A+\Gamma^-_S-\Gamma^+_A-\Gamma^+_S),
\end{equation}
where $K=(\Gamma^+_A\Gamma^-_S-\Gamma^-_A\Gamma^+_S)/(\Gamma^-_A+\Gamma^-_S+\Gamma^+_A+\Gamma^+_S)^2$, and the incoming heat from the $\mathcal{T}_A$ bath, 
\begin{equation}
Q^{A}_{\tau} =2K\tau\left[(\omega_0-\lambda)(\Gamma^+_A+\Gamma^+_S)+(\omega_0+\lambda)(\Gamma^-_A+\Gamma^-_S)\right]
\end{equation}
behave linearly with $\tau$ and efficiency reaches a plateau that depends only on the energy exchange rates and the energy levels of the system.
\begin{equation}
\eta_\tau = \frac{2 \lambda}{\omega_0}\frac{1-\frac{\Gamma^+_A+\Gamma^+_S}{\Gamma^-_A+\Gamma^-_S}}{1 +\frac{\lambda}{\omega_0}+\frac{\Gamma^+_A+\Gamma^+_S}{\Gamma^-_A+\Gamma^-_S}(1- \frac{\lambda}{\omega_0})}
\label{eta_short}
\end{equation}
For a fixed $\lambda/\omega_0$ ratio, $\eta_\tau$ is maximized when $ \Gamma^-_A+\Gamma^-_S \gg  \Gamma^+_A+\Gamma^+_S$. This can be achieved either if both $\mathcal{T}_A,\mathcal{T}_S \ll 1$ ($\Gamma^-_j \gg \Gamma^+_j)$ or if the cold reservoir ($\mathcal{T}_S$ in our case) couples to the system in a much stronger way than the hot one ($\Gamma^-_S \gg \Gamma^-_A$). In both scenarios, 
\begin{equation} 
\eta_\tau \approx 2\lambda/(\omega_0+\lambda)=1-\frac{\omega_0-\lambda}{\omega_0+\lambda}=1-\frac{E_S}{E_A},
\end{equation}
where $E_j$ is the energy of level $|j\rangle$. In the first case, the power $P_\tau =\mathcal{E}_\tau/\tau$ delivered by the machine in the short cycle depends on both temperatures and is approximately given by $P_\tau  \approx 4\lambda (\Gamma^+_A\Gamma^-_S-\Gamma^-_A\Gamma^+_S)/(\Gamma^-_A+\Gamma^-_S)$, whereas, in the second case, it depends solely on the coupling to the hot reservoir: $P_\tau \approx 4\lambda \Gamma
^+_A$. 

Finally, we show in Fig.~\ref{efficiency} that a periodic Floquet-type NESS, here referred as $\rho_{OSS}$, is indeed reached for any choice of $\tau$. Fig.~\ref{efficiency} (b) shows how many cycles are required to establish a steady operation for different periods, whereas Fig.~\ref{efficiency} (c) shows the stabilisation and the time evolution of the ergotropy of the machine for a particular choice of $\tau$.

Before concluding, let us remark that the main limitations to increase efficiency and power at short cycle operation are still the same ones for the model to hold true. On one hand, $\tau \gg \tau_d$, i.e. stage two must be much faster than stage one. On the other hand, the maximum efficiency and power require that $\sum_j \Gamma^{\pm}_j \tau \ll 1$ (the short cycle regime). That ultimately limits $\Gamma^+_A$, and, thus, the generated power. How these limitations affect the operation of the machine will depend on the specificity of its physical implementation. That said, favorable conditions are already reachable in many practical setups as it is clear in quantum jumps experiments performed in different platforms
~\cite{QuantumJump02, Haroche01, Haroche01, Devoret1, PhysRevLett.106.110502}.

We remark that the physical system described above can be implemented, as 
already discussed in a previous work~\cite{tacchino_steady_2018}, by two qubits that dissipate energy individually at rate $\gamma$ and are incoherently pumped at rate $p$, while simultaneously coupled to a common superradiant bath that dissipates energy through the symmetric decay lines at rate $\Gamma$. In this case, as long as $\lambda \ll \omega_0$, the rates become $\Gamma^+_S = \Gamma^+_A = p/2$, $\Gamma^-_A = \gamma/2$ and $\Gamma^-_S = \gamma/2+\Gamma$ and the effective temperatures are given by $\mathcal{T}_S =\frac{\omega_0}{\log\left(\frac{2\Gamma+\gamma}{p}\right)}$ and $\mathcal{T}_A = \frac{\omega_0}{\log\left(\frac{\gamma}{p}\right)}$. Maximum efficiency $\eta_\tau = \frac{2\lambda}{\omega_0+\lambda}$ is reached for $\Gamma \gg \gamma, p$, when the power becomes $P_\tau\approx 2p\lambda$.

\section{Discussion}
In summary, we have demonstrated that non-equilibrium steady states resulting from the application of a thermal gradient onto a quantum system, constitute a thermodynamic resource. We have exemplified this statement by studying a model of two coupled qubits as the working fluid. We have shown that both efficiency and output power are maximized in the short cycle limit. We notice that the out-of-equilibrium quantum thermal machine discussed in this work can be realized in a variety of quantum technological platforms nowadays. On one hand, artificial atoms, such as semiconductor quantum dots \cite{Josefsson2018} and superconducting quantum circuits \cite{Pekola2015}, realize engineered coupled qubits and represent ideal, yet not fully explored playgrounds for quantum thermodynamics demonstrations. On an alternative ground, two trapped atoms or ions naturally represent a straightforward realization of quantum working fluids \cite{Rossnagel2016}. The former solid state platforms might be preferred due to the possibility of engineering the system parameters, such as energy gaps and coupling rates, where in natural systems these parameters of the model are fixed or with limited tunability, thus producing low efficiency and output power.

We remark that the presented concept is general and widely applicable. The quantum battery need not be a two-qubit system, on the contrary it can generally be a many-body quantum system comprising e.g., several qubits, or qutrits etc.., harmonic or anharmonic oscillators, coupled via short-range or long-range forces. Accordingly, the presented concept lends itselfs to a promising broader investigation regarding the possible positive impact of many-body phenomena, such as many-body entanglement, phase transitions, collective behaviour (e.g. superradiance) on the thermodynamic performance of NESS based engines. Note that previous works have already pointed out their positive impact on the standard charging of quantum batteries by external fields, see e.g., \cite{PhysRevLett.118.150601,Ferraro18PRL120}, and on the performance of Otto engines \cite{Campisi16NATCOMM7}. Other meaningful follow up directions of the current work are related to the study of the impact of non-Markovian (instead of Lindblad-type) recharging first stages on both efficiency and generated power, as well as the effects of a non-ideal energy extracting second stage.

\section*{Acknowledgements}{MFS acknowledges FAPERJ Project No. E-26/202.576/2019 and CNPq Projects No. 302872/2019-1 and INCT-IQ 465469/2014-0. MFS would also like to thank the CICOPS program from the University of Pavia for hospitality and support. TFFS acknowledges CAPES for financial support. MC acknowledges financial support from Fondazione CR Firenze Project No. 2018.0951.}

\end{document}